\documentclass[12pt,preprint]{aastex}

\shorttitle{Large Molecules in Comets LINEAR and NEAT}
\shortauthors{Remijan et al.}
\begin{document}

\title{A BIMA ARRAY SURVEY OF MOLECULES IN COMETS LINEAR (C/2002 T7) AND NEAT (C/2001 Q4)}

\author{Anthony J. Remijan\altaffilmark{1,2,3}, D. N. Friedel\altaffilmark{4}, Imke de Pater\altaffilmark{5},
M. R. Hogerheijde\altaffilmark{6}, L. E. Snyder\altaffilmark{4}, M. F. A'Hearn\altaffilmark{7}, Geoffrey A. Blake\altaffilmark{8},
H. R. Dickel\altaffilmark{4,9}, J. R. Forster\altaffilmark{5}, C. Kraybill\altaffilmark{5}, L. W. Looney\altaffilmark{4}, Patrick Palmer\altaffilmark{10}, \& M. C. H. Wright\altaffilmark{5}}

\altaffiltext{1}{NASA Goddard Space Flight Center, Computational and Information 
Sciences and Technologies Office, Code 606, Greenbelt, MD 20771}
\altaffiltext{2}{National Research Council Resident Research Associate}
\altaffiltext{3}{Current address: National Radio Astronomy Observatory, 520 Edgemont Road, Charlottesville, VA 22901\\
email: aremijan@nrao.edu}

\altaffiltext{4}{Department of Astronomy, 1002 W. Green St., University of
Illinois, Urbana IL 61801\\
email: friedel@astro.uiuc.edu, snyder@astro.uiuc.edu, lwl@astro.uiuc.edu}

\altaffiltext{5}{Department of Astronomy, University of California, Berkeley, CA 94720\\
email: imke@floris.berkeley.edu, rforster@astro.berkeley.edu, ckraybill@astro.berkeley.edu,\\
wright@astro.berkeley.edu}

\altaffiltext{6}{Leiden Observatory, PO Box 9513, 2300 RA, Leiden, The Netherlands\\
email: michiel@strw.leidenuniv.nl}

\altaffiltext{7}{Department of Astronomy, University of Maryland, College
Park MD 20742-2421\\
email: ma@astro.umd.edu}

\altaffiltext{8}{Division of Geological and Planetary Sciences; Division of Chemistry and Chemical Engineering 
California Institute of Technology 150-21, Pasadena, CA 91125\\
email: gab@gps.caltech.edu}

\altaffiltext{9}{Current address: Department of Physics \& Astronomy, University of New Mexico,
800 Yale Blvd. NE, Albuquerque, NM 87131\\
email: h-dickel@phys.unm.edu}

\altaffiltext{10}{Department of Astronomy and Astrophysics, University of Chicago, Chicago, IL 60637\\
email: ppalmer@oskar.uchicago.edu}

\clearpage

\begin{abstract}

We present an interferometric search for large molecules, including methanol (CH$_3$OH),
methyl cyanide (CH$_3$CN), ethyl cyanide (CH$_3$CH$_2$CN), ethanol (CH$_3$CH$_2$OH), and 
methyl formate (CH$_3$OCHO), in comets LINEAR (C/2002 T7) and NEAT (C/2001 Q4) with the 
Berkeley-Illinois-Maryland Association (BIMA) array. 
In addition, we also searched for transitions of the simpler molecules
CS, SiO, HNC, HN$^{13}$C and $^{13}$CO .  We detected transitions of CH$_3$OH and CS around 
Comet LINEAR and one transition of CH$_3$OH around Comet NEAT within a synthesized 
beam of $\sim$20$''$.  We calculated the total column density and production rate of each
molecular species using the variable temperature and outflow velocity (VTOV) model
described by Friedel et al.\ (2005).
Considering the molecular production rate ratios with respect to water, Comet T7 LINEAR is
more similar to Comet Hale-Bopp while Comet Q4 NEAT is more similar to Comet Hyakutake. 
It is unclear, however, due to such a small sample size, whether there is a clear
distinction between a Hale-Bopp and Hyakutake class of comet or whether 
comets have a continuous range of molecular production rate ratios. 
\end{abstract}


\section{INTRODUCTION}

Comets are believed to contain the most pristine material remaining from the
presolar nebula and
are providing important insights into the formation mechanisms 
of complex molecular species. 
Comets are primarily located in two distinct regions of the solar
system. The Oort cloud, the source of long period (P$>$200 years) comets, is 
a spherically symmetric distribution of comets that encompasses the solar system 
out to a distance of nearly 100,000 AU. The Kuiper-Edgeworth belt that lies in the 
ecliptic plane just beyond the orbit of Pluto out to several hundred AU is the source of short 
period (20$<$P$<$200 years) comets (Jewitt 2004).
Recent models have suggested that the Oort cloud comets may have had origins in the 
entire giant planet region between Jupiter and Neptune.  These models suggest
that comets were formed in a much wider range of nebular environments and probably experienced 
thermal and collisional processing before they were ejected into the Oort cloud 
(Weissman 1999).  This processing may have ``homogenized'' the cometary nuclei of 
Oort cloud comets.  Observations comparing long and short period comets, 
which are believed to have had different origins, will allow us to address the 
molecular diversity in comets (A'Hearn et al.\ 1995, Mumma et al.\ 2005).  

Furthermore, a key goal in astrochemistry is to learn whether the 
molecular diversity seen in high and low mass hot molecular cores (HMCs) is similar to 
the chemistry of the primordial solar nebula.  The molecular diversity in comets may 
provide the link between the formation mechanisms of complex molecular species
seen in high mass HMCs and low mass young stellar objects to the formation of 
young solar systems.   Therefore, only through comprehensive surveys of HMCs and comets 
will a consistent, coherent picture emerge regarding the basic chemistry involved 
in the evolution of HMCs and comets.

Most observations of the molecular composition of cometary
gas and dust have taken place in the visible, ultraviolet (see e.g.\ Hutsem\'{e}kers et al.\ 2005 
and references therein) or infrared parts of the spectrum (see e.g.\ Helbert et al.\ 2005; 
Dello Russo et al.\ 2001).  Also, a vast inventory of interstellar 
ices and volatiles was discovered by satellites passing through the coma of Comet Halley 
(see e.g.\ Mitchell et al.\ 1992).  However, there have been 
several successful detections of the rotational transitions of molecular species, including 
organic compounds, at millimeter wavelengths (see e.g.\ Crovisier et al.\ 2004a,b; 
Biver et al.\ 2002). Many of these species are important in prebiotic organic chemistry 
(e.g.\ H$_2$O, HCN, CH$_3$OH, aldehydes, nitriles) and are believed to be parent molecules 
(originating from the comet nucleus) rather than the products of photodissociation or gas 
phase chemistry in the cometary coma.  Two prominent examples of parent molecules of prebiotic 
importance are hydrogen cyanide (HCN) (see e.g.\ Friedel et al.\ 2005) 
and methanol (CH$_{3}$OH) (see e.g.\ Ikeda et al.\ 2002).  

In the Spring of 2004, there was a rare opportunity to observe two dynamically
new Oort cloud comets passing into the inner solar system and within $\sim$0.3 AU of the Earth: Comet
LINEAR (C/2002 T7) (hereafter, T7 LINEAR) and Comet NEAT (C/2001 Q4) (hereafter, Q4 NEAT).  
This paper presents the results of an effort to further investigate the 
astrochemistry of comets by observing several large molecular species in Comets T7 
LINEAR and Q4 NEAT with the Berkeley-Illinois-Maryland Association (BIMA) array\footnote{Operated by the
University of California, Berkeley, the University of Illinois, and the
University of Maryland with support from the National Science Foundation.}; these species  
include methanol (CH$_3$OH), methyl cyanide (CH$_3$CN), ethyl cyanide (CH$_3$CH$_2$CN), 
ethanol (CH$_3$CH$_2$OH), and methyl formate (CH$_3$OCHO). 
In addition, we also searched for transitions of the simpler molecules
CS, SiO, HNC, HN$^{13}$C and $^{13}$CO.  We 
were successful in obtaining single-field images, cross-correlation spectra, 
and production rates for cometary methanol (CH$_3$OH) in both comets and 
CS in Comet T7 LINEAR.  Upper limits were found for the remaining species in both
comets.  

\section{OBSERVATIONS AND RESULTS}

The observations were conducted using the BIMA array near Hat Creek, California\footnote{121$^o$28$'$8$''$.0 
West, 40$^o$49$'$4$''$.1 North; altitude 1033 m} in D-configuration 
(baselines from $\sim$6m to $\sim$35m), cross-correlation mode toward
Comets T7 LINEAR and Q4 NEAT during their 2004 apparitions. 

The observations of Comet T7 LINEAR, using JPL reference orbit 69, were taken 2004 
May 11-15 near $\alpha(J2000)=01^h40^m$, $\delta(J2000)=-09\degr30'$\footnote{Both 
comets moved substantially across the sky during the observations, thus only approximate 
coordinates are given.}.  The comet was at a heliocentric distance of 0.73-0.77 AU 
and a geocentric distance of 0.32-0.44 AU (1\arcsec=319 km at 0.44 AU).
W3(OH) was used as the flux density calibrator for these observations. The quasar
0108+015 was used to calibrate the antenna based gains.  The observations of Comet Q4
NEAT, using JPL reference orbit 123, were taken 2004 May 20-24 near 
$\alpha(J2000)=09^h07^m$, $\delta(J2000)=34\degr30'$\footnotemark[11].
The comet was at a heliocentric distance of $\sim$0.97 AU and a geocentric distance
of 0.55-0.61 AU (1\arcsec=442 km at 0.61 AU).  Mars was used as the flux density 
calibrator and 0927+390 was used to calibrate the antenna based gains. The absolute amplitude 
calibration is accurate to within $\sim$20\%.  The channel spacing of all observations 
was 0.391 MHz, except for CS which had a spacing of 0.098 MHz.  The data were reduced, combined, 
and imaged using the MIRIAD software package (Sault et al.\ 1995).

Table 1 lists the molecular line parameters of the searched species in Comets
T7 LINEAR and Q4 NEAT.  The first column lists the molecular species, the second lists the transition, the 
third lists the transition frequency (MHz), the fourth lists the upper state energy level
of the transition (K), the fifth lists the line strength multiplied by the square of the
electronic dipole moment (D$^2$), and the last column lists the 
quiet sun photodissociation rate for each species (s$^{-1}$). Table 2 lists the observational 
parameters. For each comet, the heliocentric distance ($r_H$), geocentric distance 
($\Delta$) in astronomical units (AU), and the synthesized beam size ($''$$\times$$''$) are listed. 

Figure 1(a-c) display the map and spectra of CH$_3$OH around Comet T7 LINEAR.
Figure 1a shows the contour image of the $3_{1,3}-4_{0,4} A+$ transition of CH$_3$OH starting 
at 3 $\sigma$.  
Figure 1b shows the cross-correlation spectrum of this transition. The dashed line 
corresponds to the rest frequency of the $3_{1,3}-4_{0,4} A+$ line for a cometocentric rest 
velocity of 0 km s$^{-1}$. The 1 $\sigma$ rms noise level is indicated at the left of the panel. 
The CH$_3$OH line for Comet T7 LINEAR was fit with a Gaussian by a least-squares method which 
gives a peak intensity of 0.44(0.06) Jy beam$^{-1}$ and a FWHM of 1.74(0.31) km s$^{-1}$.
Figure 1c shows the cross-correlation spectrum (Hanning smoothed over three channels) of this transition.
Figure 1(d-f) display the map and spectra of CH$_3$OH around Comet Q4 NEAT.
Figure 1d shows the contour image of CH$_3$OH emission starting at 3 $\sigma$.
Figure 1e is the cross-correlation spectrum and the dashed line is similar to Figure 1b for a 
cometocentric rest velocity of 0 km s$^{-1}$. The least-squares Gaussian line fit for Comet Q4 
NEAT gives a peak intensity of 0.31(0.13) Jy beam$^{-1}$ and a FWHM of 0.72(0.3) km s$^{-1}$.  
Figure 1f is the Hanning smoothed cross-correlation spectrum.

Figure 2(a-c) display the map and spectra of CS around Comet T7 LINEAR. Figure 2a shows the contour image of the 
$J=2-1$ transition of CS starting at 2 $\sigma$. 
Figure 2b is the cross-correlation spectrum 
of this transition. The least-squares Gaussian line fit for Comet T7 LINEAR gives a peak intensity
of 0.21(0.05) Jy beam$^{-1}$ and a FWHM of 2.45(0.60) km s$^{-1}$. Figure 2c is the Hanning
smoothed cross-correlation spectrum. 

All other molecular transitions in Table 1, including CS in Comet Q4 NEAT, were not detected above 3 
$\sigma$. Thus, all column density and production rate upper limits were calculated using the 
1 $\sigma$ rms noise level of the window containing the molecular line emission
and a line width of one channel ($\sim$1.3 km s$^{-1}$). Tables 3 and 4 list 
either the measured intensities (Jy beam$^{-1}$) and line widths (km s$^{-1}$) or the upper limits of the intensity 
and line width for each species in columns 2 and 3.  Column 4 lists the total beam-averaged 
molecular column density (cm$^{-2}$) and column 5, the production rate (s$^{-1}$).  The 
calculated production rates were determined using the variable temperature and outflow velocity (VTOV) 
model (Friedel et al.\ 2005).  Finally, columns 6 and 7 list the production rate ratios with respect to 
H$_2$O and CN as will be discussed in $\S$3.2.

\section{DISCUSSION}
\subsection{Column Densities and Production Rates}\label{sec:NT}

As described in Friedel et al.\ (2005), the VTOV model calculates
the total column density and production rate of a cometary species assuming optically thin 
emission in LTE and that the temperature and outflow velocity within the coma vary with 
cometocentric distance.  The VTOV model calculations to determine the total column density 
and production rate are discussed in detail in Friedel et al.\ (2005) Appendix A. 
Here, we discuss the implication of the production rates and column densities. 
Using the VTOV model, tables 3 and 4 list the total beam-averaged column densities and 
production rates of each molecular species observed toward Comets T7 LINEAR and Q4 NEAT.
The VTOV model was also used to determine the total beam-averaged column density and 
production rate of CS even though it is believed to be the daughter species.
The parent molecule of CS may be CS$_2$, which has a very short lifetime of $\sim$10$^3$ s 
or less (Snyder et al.\ 2001).  This short lifetime suggests that CS is formed in the inner 
coma and the measured line width of CS in Comet LINEAR is similar to HCN and CH$_3$OH, which 
are known parent species. Thus, at the spatial resolution of our observations, CS can be fitted as 
though it were a parent species.


\subsection{Relative Production Rates of X to H$_2$O and CN}

\subsubsection{Comet LINEAR}
Based upon the H$_2$O production rates as measured by Schleicher et al.\ (2005, private communication) 
at different heliocentric distances, Friedel et al.\ (2005)  determined the H$_2$O production rate to 
be $\sim2\times10^{29}$ s$^{-1}$
during our observing period.  The relative production rates of each molecular species relative to H$_2$O are 
listed in column 6 of Table 3.
The relative production rate ratios of HCN (Friedel et al.\ 2005), CH$_3$OH, and CS with respect to H$_2$O 
for Comet LINEAR are most similar to those observed around Comet Hale-Bopp (C/1995 O1) 
(($2.1-2.6)\times10^{-3}$, $4\times10^{-2}$, and $(3.7-4.3)\times10^{-3}$, respectively) (Snyder et al.\ 2001,
Remijan et al.\ 2004). Schleicher et al.\ (2005) estimated the CN production rate to be 
$\sim1.4\times10^{26}$ s$^{-1}$.  We also attempted to determine the relative production rates of each 
nitrogen bearing molecular species relative to CN. However in this work, no species observed toward Comet 
LINEAR with a CN bond was detected above the 3 $\sigma$ detection limit.  The upper limits to the production 
rate ratios are listed in column 7 of Table 3.

\subsubsection{Comet NEAT}

Schleicher et al.\ (2005) also determined the H$_2$O and CN production rates for comet NEAT during our 
observing period (as reported by Friedel et al.\ 2005).  These are
very similar to water production rates measured from Comet Hyakutake at similar heliocentric distances  
($\sim1\times10^{29}$ s$^{-1}$ Lis et al.\ 1997).  The relative production rates of each 
molecular species relative to H$_2$O are listed in column 6 of Table 4.  The relative production rate ratios 
of HCN (Friedel et al.\ 2005) and CH$_3$OH relative to H$_2$O are most similar to those measured around Comet 
Hyakutake (C/1996 B2) ($1.2\times10^{-3}$ and $1.7\times10^{-2}$, respectively) (Biver et al.\ 1999).
The average CN production rate is Q(CN)$\sim2.6\times10^{26}$ s$^{-1}$.  The relative production rates 
of each nitrogen bearing molecular species relative to CN are listed in column 7 of Table 4, however
the only species observed toward Comet NEAT with a CN bond was CH$_3$CN which was not detected above
the 3 $\sigma$ detection limit.  The upper limit to the production rate ratio was $<$4.0$\times$10$^{-2}$.


\subsection{Classifying Comets T7 LINEAR and Q4 NEAT}

Comets are diverse objects, as has been noted by many researchers in the past. For example, A'Hearn et al.\ 
(1995) reported an optical study of 85 comets, Biver et al.\ 
(2002) reached this conclusion based upon a radio survey of 24 comets, while Mumma et al.\ (2005) reached 
the same conclusion from the hypervolatile gases CO and CH$_4$ from observations at IR wavelengths. Many comet 
observers have searched for differences between comets from the Oort cloud and the short-period (Jupiter family) 
comets to attempt to find clear differences in chemical composition depending on their place of formation. 
As more observations become available, it becomes harder to differentiate between the various comets classes, 
in particular after the Deep Impact mission (e.g., Mumma et al, 2005). 

We can now include Comets T7 LINEAR and Q4 NEAT
in the chemical comparison from the detections of both CH$_3$OH and HCN (Friedel et
al.\ 2005).  Both comets T7 LINEAR and Q4 NEAT are believed to be Oort 
cloud comets given their measured orbital parameters and our observations show that both
comets are chemically very similar.  However, from the Biver et al.\ (2002) 30 m 
survey, there appears to be a distinction between comets rich in HCN 
(HCN/H$_2$O$>$0.2\%) including comets Hale-Bopp (C/1995 01), 109P/Swift-Tuttle, 
1P/Halley and 9P/Tempel 1 (Mumma et al.\ 2005) compared to  
comets with an HCN/H$_2$O abundance ratio $\sim$0.1\% which include, among several 
others, comets Hyakutake (C/1996 B2), Austin (C/1989 X1) and Levy (C/1990 K1).  
It is unclear, however, due to such a small sample size, whether in fact there are two distinct 
classes of comets (i.e.\ Hale-Bopp and Hyakutake class) or if comets have a range of different 
molecular production rate ratios. Given the abundance ratios of Comet T7 LINEAR, it is more similar Comet
Hale-Bopp that was both HCN and CH$_3$OH rich.  Comet Q4 NEAT on the
other hand, has column densities of HCN and CH$_3$OH more similar to Comet Hyakutake.  
Furthermore, this distinction appears across both Oort cloud 
and Kuiper-Edgeworth belt comets as illustrated by recent Deep Impact observations
of Comet 9P/Tempel 1 (Mumma et al.\ 2005).  
Thus, while our observations still tend to support a chemical distinction between Hale-Bopp 
and Hyakutake class comets due to the abundance rations of CH$_3$OH/H$_2$O and HCN/H$_2$O,
more observations of both long and short period comets are necessary to increase the 
sample size.  With better statistics, it may be possible to reliably determine the origins 
of both Oort cloud and Kuiper-Edgeworth belt comets in the pre-solar nebula.

\section{SUMMARY}

In this paper, we have presented the results of an interferometric search for several large 
molecules including, methanol (CH$_3$OH), methyl cyanide (CH$_3$CN), ethyl cyanide (CH$_3$CH$_2$CN), 
ethanol (CH$_3$CH$_2$OH), and methyl formate (CH$_3$OCHO) in 
Comets T7 LINEAR and Q4 NEAT. In addition, we also searched for transitions of the simpler molecules
CS, SiO, HNC, HN$^{13}$C and $^{13}$CO.  Of the 10 species listed in Table 1, we detected transitions of CH$_3$OH 
and CS around Comet T7 LINEAR and one transition of CH$_3$OH around Comet Q4 NEAT.
Using the variable temperature and outflow velocity (VTOV) model described by Friedel 
et al.\ (2005), we determined the total beam-averaged column densities and production 
rates of each molecular species observed toward Comets T7 LINEAR and Q4 NEAT.  Based on the 
molecular production rate ratios with respect to water it appears 
that Comet T7 LINEAR is more similar to Comet Hale-Bopp while Comet Q4 NEAT is more 
similar to Comet Hyakutake.  However, due to such a small sample size, it is unclear 
whether there are two distinct classes of comet or if there is a continuous range of classes 
with numerous and different molecular production rate ratios.  More observations with
higher sensitivity interferometers such as CARMA or ALMA of both long 
and short period comets are necessary to increase the sample size, hopefully allowing a determination of the 
origins of both Oort cloud and Kuiper-Edgeworth belt comets in the pre-solar nebula.

\acknowledgements
We thank J.~R. Dickel for assisting with the observations, an anonymous referee for many helpful comments,
and D.~G. Schleicher for providing  H$_2$O and CN production rates. 
This work was partially funded by: NSF AST02-28953, AST02-28963, AST02-28974 and AST02-28955; 
and the Universities of Illinois, Maryland, and California, Berkeley.

\clearpage
\begin{deluxetable}{lccccc}
\tablewidth{0pt}
\tablecolumns{6}
\tablecaption{Molecular Line Parameters}
\tablehead{
\colhead{\small } &\colhead{\small Quantum} & \colhead{\small Frequency} & \colhead{\small $E_{u}$} & \colhead{\small $S\mu^{2}$} & \colhead{$\alpha$(\small 1 AU)\tablenotemark{a}}\\
\colhead{\small Species} &\colhead{\small Numbers} & \colhead{\small (MHz)} & \colhead{\small (K)} & \colhead{\small (D$^{2}$)} & \colhead{\small (s$^{-1}$)}
}
\startdata
{\small CH$_3$OH }&{\small  $3_{1,3}-4_{0,4}A+$ }&{\small  107,013.770(13) }&{\small  28.3 }&{\small  3.1 }&{\small  $1.3\times10^{-5}$ }\\
{\small CS }&{\small  $2-1$ }&{\small  97,980.985(15) }&{\small  7.7 }&{\small  7.1 }&{\small  $6.7\times10^{-6}$ }\\
{\small $^{13}$CO }&{\small  $1-0$ }&{\small  110,201.353(1) }&{\small  5.3 }&{\small  0.01 }&{\small  $1.2\times10^{-6}$\tablenotemark{b}}\\
{\small CH$_3$CN }&{\small  $6_K-5_K$ }&{\small  110,383.502(2)\tablenotemark{c} }&{\small  18.5 }&{\small  91.9 }&{\small  $2.9\times10^{-6}$ }\\
{\small CH$_3$CH$_2$CN }&{\small  $10_{1,10}-9_{1,9}$ }&{\small  86,819.846(0) }&{\small  24.1 }&{\small  146.7 }&{\small  $\sim10^{-5}$ }\\
{\small SiO }&{\small  $2-1$ }&{\small  86,846.960(50) }&{\small  6.3 }&{\small  19.2 }&{\small  $6.7\times10^{-6}$}\\
{\small HN$^{13}$C}&{\small  $1-0$ }&{\small  87,090.850(50) }&{\small  4.2 }&{\small  7.3 }&{\small  $1.5\times10^{-5}$\tablenotemark{d}}\\
{\small CH$_3$CH$_2$OH }&{\small  $4_{1,4}-3_{0,3}$ }&{\small  90,117.576(2) }&{\small  9.4 }&{\small  5.4 }&{\small  $1.8\times10^{-5}$}\\
{\small CH$_3$OCHO }&{\small  $8_{0,8}-7_{0,7}E$ }&{\small  90,227.595(13) }&{\small  20.1 }&{\small  21.0 }&{\small  $4.7\times10^{-5}$ }\\
{\small }&{\small  $8_{0,8}-7_{0,7}A$ }&{\small  90,229.647(14) }&{\small  20.1 }&{\small  21.0 }&{\small  $4.7\times10^{-5}$}\\
{\small HNC }&{\small  $1-0$ }&{\small  90,663.572(4) }&{\small  4.4 }&{\small  9.3 }&{\small  $1.5\times10^{-5}$}\\
\enddata
\tablecomments{\footnotesize All molecular line parameters taken from Pickett et al. (1998).  Errors in the rest frequency are
2 $\sigma$.}
\tablenotetext{\footnotesize a}{\footnotesize Taken or estimated from Crovisier et al.\ (1994).}
\tablenotetext{\footnotesize b}{\footnotesize Assumed to be the same as for CO.}
\tablenotetext{\footnotesize c}{\footnotesize Only the frequency of the $K=0$ transition is given.}
\tablenotetext{\footnotesize d}{\footnotesize Assumed to be the same as for HNC.}
\end{deluxetable}

\begin{deluxetable}{lcccccccc}
\tablewidth{0pt}
\tablecolumns{8}
\tablecaption{Observational Parameters}
\tablehead{
\colhead{\small } & \multicolumn{3}{c}{\small LINEAR} & \colhead{} & \multicolumn{3}{c}{\small NEAT} \\
\cline{2-4}\cline{6-8}\colhead{\small } & \colhead{\small $r_H$} & \colhead{\small $\Delta$} & \colhead{\small $\theta_a\times\theta_b$} & \colhead{\small } & \colhead{\small $r_H$} & \colhead{\small $\Delta$} & \colhead{\small $\theta_a\times\theta_b$}\\
\colhead{\small Species} & \colhead{\small (AU)} & \colhead{\small (AU)} & \colhead{\small (\arcsec$\times$\arcsec)} & \colhead{\small } & \colhead{\small (AU)} & \colhead{\small (AU)} & \colhead{\small (\arcsec$\times$\arcsec)}
}
\startdata
{\small  CH$_3$OH }&{\small  0.75 }&{\small  0.38 }&{\small  $22.3\times16.9$ }&{\small  }&{\small  0.97 }&{\small  0.55 }&{\small  $16.7\times14.3$ }\\
{\small  CS }&{\small  0.74 }&{\small  0.41 }&{\small  $23.7\times18.9$ }&{\small  }&{\small  0.97 }&{\small  0.55 }&{\small  $18.8\times15.4$ }\\
{\small  $^{13}$CO }&{\small  0.75 }&{\small  0.38 }&{\small  $21.8\times16.1$ }&{\small  }&{\small  0.97 }&{\small  0.55 }&{\small  $16.0\times14.1$ }\\
{\small  CH$_3$CN }&{\small  0.75 }&{\small  0.38 }&{\small  $21.6\times16.4$ }&{\small  }&{\small  0.97 }&{\small  0.55 }&{\small  $16.2\times14.0$ }\\
{\small  CH$_3$CH$_2$CN }&{\small  0.77 }&{\small  0.32 }&{\small  $28.5\times21.3$ }&{\small  }&{\small  }&{\small  }&{\small  }\\
{\small  SiO }&{\small  0.77 }&{\small  0.32 }&{\small  $28.5\times21.3$ }&{\small  }&{\small  }&{\small  }&{\small  }\\
{\small  HN$^{13}$C }&{\small  0.77 }&{\small  0.32 }&{\small  $28.7\times20.0$ }&{\small  }&{\small  }&{\small  }&{\small  }\\
{\small  CH$_3$CH$_2$OH }&{\small  0.77 }&{\small  0.32 }&{\small  $28.2\times19.6$ }&{\small  }&{\small  }&{\small  }&{\small  }\\
{\small  CH$_3$OCHO }&{\small  0.77 }&{\small  0.32 }&{\small  $27.7\times19.3$ }&{\small  }&{\small  }&{\small  }&{\small  }\\
{\small  HNC }&{\small  0.77 }&{\small  0.32 }&{\small  $27.7\times20.4$ }&{\small  }&{\small  }&{\small  }&{\small  }\\
\enddata
\end{deluxetable}

\begin{deluxetable}{lcccccc}
\tablewidth{0pt}
\tablecolumns{7}
\tablecaption{Variable Temperature and Outflow Velocity (VTOV) Results for Comet LINEAR}
\tablehead{
\colhead{\small } & \colhead{\small $I_0$} & \colhead{\small $\Delta v$} &\colhead{\small $\langle N_T\rangle$} & \colhead{\small Q} & \colhead{\small } & \colhead{\small }\\
\colhead{\small Species} &\colhead{\small (Jy bm$^{-1}$)} & \colhead{\small km s$^{-1}$)} & \colhead{\small (cm$^{-2}$)} & \colhead{\small (s$^{-1}$)} & \colhead{\small $\frac{\rm Q(X)}{\rm Q(H_2O)}$} & \colhead{\small $\frac{\rm Q(X)}{\rm Q(CN)}$\tablenotemark{a}}}
\startdata
{\small CH$_3$OH }&{\small  0.44(0.06) }&{\small  1.74(0.31) }&{\small  $1.4(0.3)\times10^{14}$ }&{\small  $7.5(1.5)\times10^{27}$ }&{\small  $3.8(0.8)\times10^{-2}$ }&{\small  }\\
{\small CS }&{\small  0.21(0.05) }&{\small  2.45(0.60) }&{\small  $1.4(0.4)\times10^{12}$ }&{\small  $1.5(0.5)\times10^{27}$ }&{\small  $7.5(25)\times10^{-3}$ }&{\small  }\\
{\small CH$_3$CN }&{\small  $<0.13$ }&{\small  1.06 }&{\small  $<4.6\times10^{11}$ }&{\small  $<2.3\times10^{25}$ }&{\small  $<1.2\times10^{-4}$ }&{\small  $<0.16$}\\
{\small $^{13}$CO }&{\small  $<0.12$ }&{\small  1.06 }&{\small  $<1.5\times10^{14}$ }&{\small  $<7.7\times10^{27}$ }&{\small  $<4.0\times10^{-2}$ }&{\small  }\\
{\small CH$_3$OCHO }&{\small  $<0.09$ }&{\small  1.30 }&{\small  $<2.5\times10^{13}$ }&{\small  $<2.2\times10^{27}$ }&{\small  $<1.0\times10^{-2}$ }&{\small  }\\
{\small HNC }&{\small  $<0.08$ }&{\small  1.35 }&{\small  $<1.8\times10^{11}$ }&{\small  $<1.4\times10^{25}$ }&{\small  $<7.5\times10^{-5}$ }&{\small  $<0.10$}\\
{\small HN$^{13}$C }&{\small  $<0.08$ }&{\small  1.35 }&{\small  $<2.8\times10^{11}$ }&{\small  $<2.2\times10^{25}$ }&{\small  $<1.1\times10^{-4}$ }&{\small  $<0.16$}\\
{\small SiO }&{\small  $<0.08$ }&{\small  1.35 }&{\small  $<2.2\times10^{11}$ }&{\small  $<1.7\times10^{25}$ }&{\small  $<8.5\times10^{-5}$ }&{\small  }\\
{\small CH$_3$CH$_2$CN }&{\small  $<0.08$ }&{\small  1.30 }&{\small  $<2.3\times10^{12}$ }&{\small  $<1.8\times10^{26}$ }&{\small  $<9.0\times10^{-4}$ }&{\small  $<1.3$}\\
{\small CH$_3$CH$_2$OH }&{\small  $<0.08$ }&{\small  1.35 }&{\small  $<1.9\times10^{13}$ }&{\small  $<1.4\times10^{27}$ }&{\small  $<7.0\times10^{-3}$ }&{\small  }\\
\enddata
\tablenotetext{\footnotesize a}{\footnotesize The ratio is given only for those species which contain a C-N bond.}
\end{deluxetable}

\begin{deluxetable}{lcccccc}
\tablewidth{0pt}
\tablecolumns{7}
\tablecaption{Variable Temperature and Outflow Velocity (VTOV) Results for Comet NEAT}
\tablehead{
\colhead{\small } & \colhead{\small $I_0$} & \colhead{\small $\Delta v$} &\colhead{\small $\langle N_T\rangle$} & \colhead{\small Q} & \colhead{\small } & \colhead{\small }\\
\colhead{\small Species} &\colhead{\small (Jy bm$^{-1}$)} & \colhead{\small km s$^{-1}$)} & \colhead{\small (cm$^{-2}$)} & \colhead{\small (s$^{-1}$)} & \colhead{\small $\frac{\rm Q(X)}{\rm Q(H_2O)}$} & \colhead{\small $\frac{\rm Q(X)}{\rm Q(CN)}$\tablenotemark{a}}}
\startdata
{\small CH$_3$OH }&{\small  0.31(0.13) }&{\small  0.72(0.30) }&{\small  $4.6(2.7)\times10^{13}$ }&{\small  $2.9(1.7)\times10^{27}$ }&{\small  $2.4(1.4)\times10^{-2}$ }&{\small  }\\
{\small CS }&{\small  $<0.14$ }&{\small  0.60 }&{\small  $<4.0\times10^{11}$ }&{\small  $<1.4\times10^{25}$ }&{\small  $<1.4\times10^{-3}$ }&{\small  }\\
{\small CH$_3$CN }&{\small  $<0.06$ }&{\small  1.06 }&{\small  $<2.0\times10^{11}$ }&{\small  $<1.1\times10^{25}$ }&{\small  $<9.2\times10^{-5}$ }&{\small  $<4.0\times10^{-2}$}\\
{\small $^{13}$CO }&{\small  $<0.06$ }&{\small  1.06 }&{\small  $<9.8\times10^{13}$ }&{\small  $<8.3\times10^{27}$ }&{\small  $<7.0\times10^{-2}$ }&{\small  }\\
\enddata
\tablenotetext{\footnotesize a}{\footnotesize The ratio is given only for those species which contain a C-N bond.}
\end{deluxetable}

\clearpage
\figcaption{Comet LINEAR (C/2002 T7) and NEAT (C/2001 Q4) single field CH$_3$OH images and spectra. (a) Comet T7 emission 
contours from the $J=3_{1,3}-4_{0,4} A+$ transition of CH$_3$OH at 107.013 GHz. The contour levels are -0.2, 0.3, 0.4, 0.5, 
and 0.6 Jy/beam ($1 \sigma$ spacing, starting at 3 $\sigma$). Image coordinates are arcseconds offsets relative to the 
predicted position of the nucleus. The synthesized beam is in the lower left, and the line segment shows the solar direction. 
(b) CH$_3$OH cross-correlation spectra, Ordinate is flux density per beam, I$_\nu$, in Jy/beam; $\sigma\sim$ 0.1 Jy/beam 
(c) CH$_3$OH cross-correlation spectra (Hanning smoothed over 3 channels), labels are the same as in (b). (d) Comet Q4 emission 
contours from CH$_3$OH. Contours indicate the CH$_3$OH emission near its peak centered at a cometocentric velocity of 0 km/s. 
The contour levels are -0.1, 0.15, 0.2, and 0.25 Jy/beam ($1 \sigma$ spacing, starting at 3 $\sigma$).Image coordinates and 
labels are the same as in (a). (e) CH$_3$OH cross-correlation spectra; labels are the same as in (b), $\sigma\sim$ 0.05 
Jy/beam. (f) CH$_3$OH cross-correlation spectra (Hanning smoothed over 3 channels), abscissa and ordinate are the same as in 
(b).} 

\figcaption{Comet LINEAR (C/2002 T7) single field CS image and spectra. (a) Comet T7 emission contours from the $J=2-1$ 
transition of CS at 97.980 GHz. Contours indicate the CS emission near its peak centered at a cometocentric velocity of 0 
km/s. The contour levels are -0.15, 0.15, 0.225, and 0.3 Jy/beam ($1 \sigma$ spacing, starting at 2 $\sigma$) Image coordinates 
and labels are the same as in Figure 1. (b) CS cross-correlation spectra; labels are the same as in Figure 1(b), $\sigma\sim$ 
0.075 Jy/beam. (c) CS cross-correlation spectra (Hanning smoothed over 3 channels); labels are the same as in (b).}

\clearpage
\begin{figure}
\epsscale{0.8}
\plotone{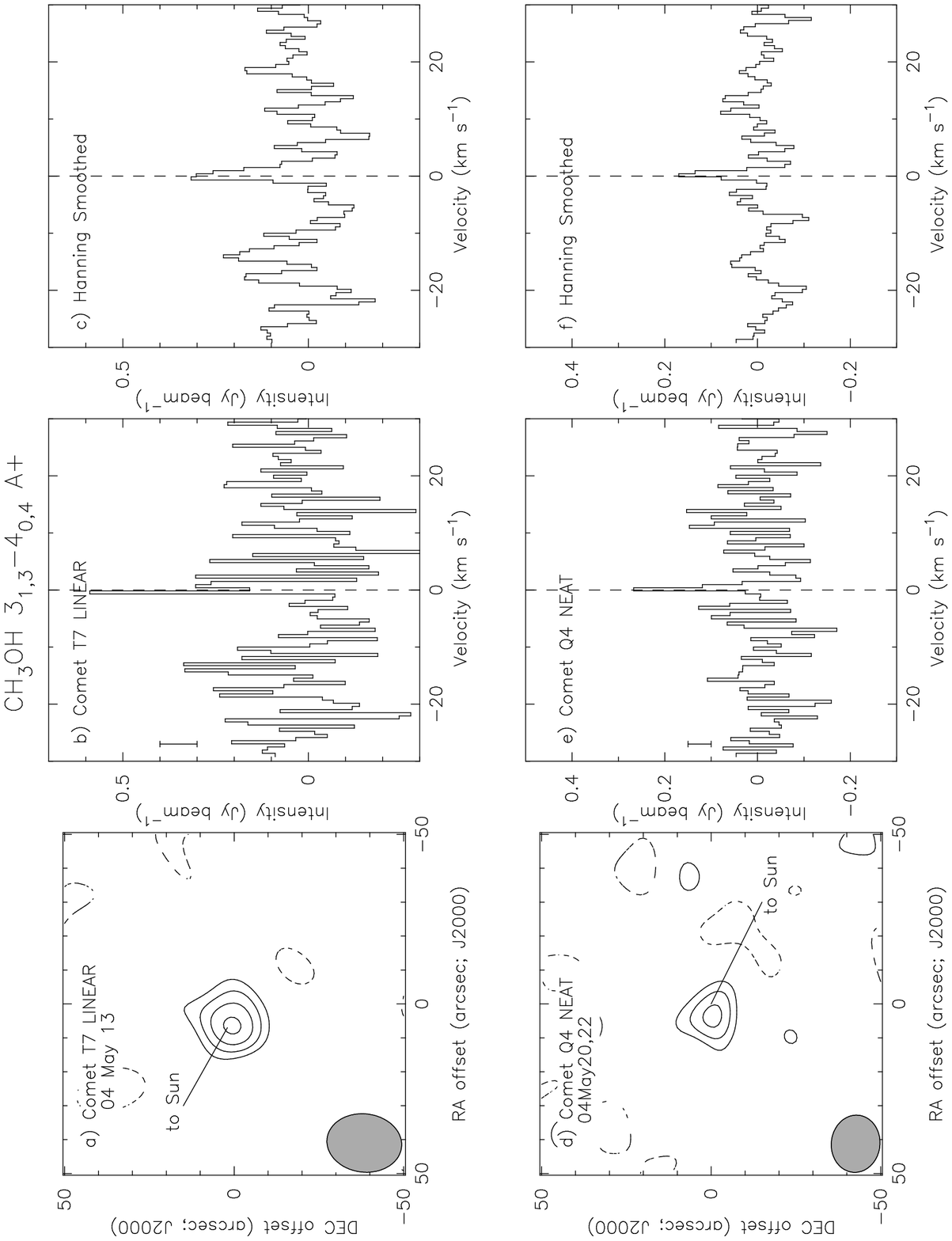}
\centerline{Figure 1.}
\end{figure}
\clearpage
\begin{figure}
\epsscale{0.45}
\plotone{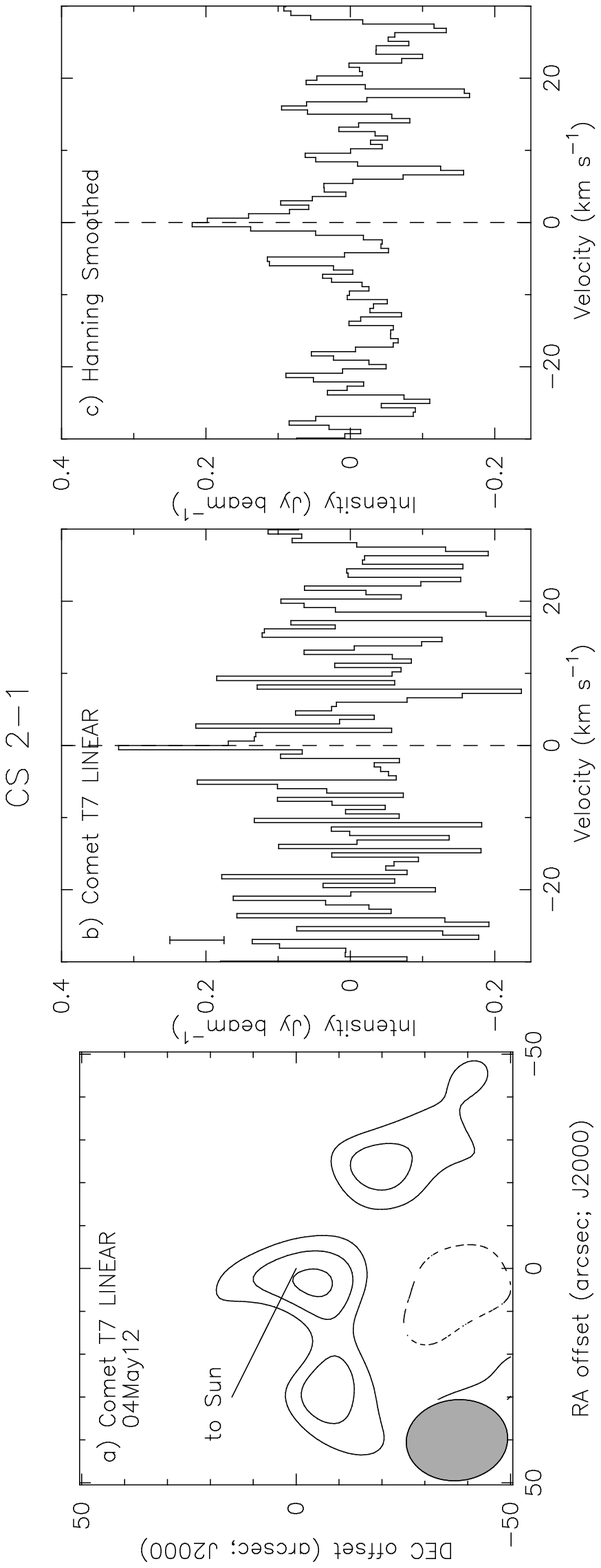}
\centerline{Figure 2.}
\end{figure}

\end{document}